# Breathing ferroelectricity induced topological valley states in kagome niobium halide monolayers

Kai-Qi Wang[1], Jun-Ding Zheng[1], Wen-Yi Tong[1, *], Chun-Gang Duan[1,2,3,*]

[1] Key Laboratory of Polar Materials and Devices, Ministry of Education, East China Normal University, Shanghai 200241, China.

[2] Collaborative Innovation Center of Extreme Optics, Shanxi University, Taiyuan, Shanxi 030006, China.

[3] Shanghai Center of Brain-inspired Intelligent Materials and Devices, East China Normal University, Shanghai 200241, China.

\* Author to whom any correspondence should be addressed.

**E-mail:** wytong@ee.ecnu.edu.cn; cgduan@clpm.ecnu.edu.cn

# ABSTRACT

In recent years, kagome lattices have garnered significant attention for their diverse properties in topology, magnetism, and electron correlations. However, the exploration of breathing kagome lattices, which exhibit dynamic breathing behavior, remains relatively scarce. Structural breathing introduces an additional degree of freedom that is anticipated to fine-tune the electronic structure, potentially leading to exotic properties within the system. In this study, we employ a combination of the $k \cdot p$ model and first-principles calculations to explore how breathing ferroelectricity can modulate valley states within a monolayer of niobium halide with breathing kagome lattice. Through the interplay of magnetoelectric coupling and the lock-in between breathing and ferroelectric states, we demonstrate that a dynamically breathing process, when controlled by an appropriately applied electric field, can achieve valley polarization reversal and generate multiple valley states, including those that are topologically nontrivial. These state transformations may couple to distinctive properties in circularly-polarized optical responses and various valley Hall effects. Consequently, our results suggest that materials featuring breathing kagome lattices represent promising platforms for studying the interplay among structure, charge, spin, and valley degrees of freedom, a crucial step toward developing multifunctional devices.

# INTRODUCTION

Kagome lattice materials, made up of corner-sharing triangles, exhibit a rich variety of phenomena that have garnered significant attention, including unconventional superconductivity[1, 2], charge density waves[3, 4], complex magnetism[5, 6], spin and pair density waves[7-9], and electronic nematicity[10]. Beyond the conventional kagome lattice, the breathing kagome lattice introduces a dynamic "breathing" behavior, where alternating triangles exhibit varying bond lengths between constituent ions, leading to different sizes. A series of materials, including $W_3Cl_8$[11], $Nd_3BWO_9$[12] and $ScOs_2$[13], belong to this family. Although these materials exhibit a wealth of interesting properties such as magneto-optical Kerr effect[11] and proximate spin liquid[12], the switching of this breathing character and its impact on other properties remain scarcely explored. We anticipate that structural breathing, which naturally breaks spatial inversion symmetry and thereby significantly influences electronic properties, can serve as an additional degree of freedom, offering the possibility to tune various performance.

On the other side, valley physics is an emerging frontier in condensed matter physics that explores the electronic behaviors associated with local minima in the conduction and/or valence bands[14, 15]. These valleys represent energetically degenerate but inequivalent momentum states in reciprocal space and offer potential in data encoding[16, 17], photodetection[18, 19] and quantum computing[20]. A key challenge in harnessing their further application for robust manipulation lies in breaking the energy degeneracy between valleys. To address this issue, the concept of ferrovalley material was proposed in 2016, enabling the use of spontaneous valley polarization (VP) for non-volatile applications[21]. Subsequently, the half-valley metal[22], as another member of the ferrovalley family, has been introduced, combining two hot topics—valleytronics and topology, together. To date, achieving the topological valley states has predominantly relied on methods such as strong correlation[22, 23] and biaxial strain[24], which remain conceptually intriguing but experimentally challenging.

It is then straightforward to explore valley physics in kagome system. Indeed, the $Nb_3X_8$ (X=Cl, Br, I), as a representative breathing kagome family, has been found to exhibit a variety of fascinating properties, such as Mott physics and correlated states[25, 26], topological flat bands[27, 28] and the magnetic-nonmagnetic phase transition[29]. More interestingly, monolayer $Nb_3I_8$ possesses ferrovalley properties and integrates ferromagnetic and ferroelectric characteristics, identifying it as a triferroic material[30, 31]. Furthermore, valley manipulation has been proposed in bilayer $Nb_3I_8$ systems through various configurations of magnetic moments and electric polarization[31, 32]. Nonetheless, achieving VP or even triferroicity requires the magnetic easy axis to be oriented out-of-plane, while its intrinsic nature is in-plane, suggesting that it is not intrinsically a ferrovalley material. In this regard, two critical questions naturally arise: how can VP be achieved in the $Nb_3I_8$, and is there any coupling between the breathing degree of freedom and valley physics?

In this work, we propose the magnetoelectric coupling mechanism to realize VP in this kind of breathing kagome lattice. Additionally, based on the unique interaction among structural breathing, ferroelectricity and valley, low-power electric methods are expected to be utilized for achieving the control of VP reversal and multiple valley states including topological ones. Combining the $k \cdot p$ model with the first-principles calculations, we use the breathing kagome $Nb_3I_8$ monolayer as an example to demonstrate that vertical electric field is able to tune the magnetic anisotropy, which serves as a new mechanism for inducing VP. Furthermore, the coupling of breathing dynamics with ferroelectricity, together with the connection between magnetoelectric effects and valley, suggests the potential for modulating VP reversal and multiple valley states via breathing ferroelectricity. During the breathing process, transitions occur among topologically trivial paravalley states with the conventional valley Hall effect (VHE), VP states with the anomalous valley Hall effect (AVHE), and nontrivial half-valley metal states with the quantum anomalous valley Hall effect (QAVHE). These findings indicate rich interactions in breathing kagome lattice, including magnetoelectric coupling and the interplay between breathing dynamics and ferroelectricity, which together give rise to reversal of VP and the multiple valley states corresponding to distinct optical absorption properties. This system can thus serve as an ideal platform for investigating the coupling among lattice, charge, spin, valley degrees of freedom, facilitate the integration of valleytronics, topology, and optics, and pave the way for practical applications in valleytronic, optoelectronic and quantum devices.

**RESULTS AND DISCUSSIONS**

The corresponding properties are studied in detail using $Nb_3I_8$ as an example. As illustrated in Figure 1(a), bulk $Nb_3I_8$ consists of six layers stacked through interlayer van der Waals stacking, exhibiting spatial inversion symmetry[33]. Meanwhile, it has two distinct interlayer spacings of 3.28 Å and 2.64 Å, corresponding to cleavage energies of 0.24 and 0.26 $J/m^2$, respectively (in Figure S1). These are lower than graphene (0.37 $J/m^2$)[34] and SnSe (0.57 $J/m^2$)[35], suggesting the possibility of exfoliation into monolayers. Furthermore, our molecular dynamics simulations and phonon calculation in Figure S4 confirm the thermodynamic and dynamical stability of monolayer $Nb_3I_8$. As expected, the $Nb_3I_8$ flakes in experiments were reduced to monolayer by the mechanical exfoliation method[36], supporting $Nb_3I_8$ as a novel candidate for further investigation of monolayer physics.

Monolayer $Nb_3I_8$, as shown in Figure 1(b), has *P*3*m*1 (No. 156) symmetry (the corresponding table of characters can be found in Table S2). The relaxed lattice constant is provided in Table S1, consistent with other works[32, 33]. A top view reveals that three edge-sharing distorted $NbI_6$ octahedra form Nb trimer clusters. The Nb-Nb distances are slightly shorter than the inter-cluster distances. The alternating triangles in the kagome lattice have different bond lengths, resulting in a breathing characteristic. In the side view, these Nb atoms are situated between two distinct layers of I atoms, allowing monolayer $Nb_3I_8$ can be regarded as a Janus structure. Contrary to common ones, its unique feature is that both the upper and lower layers consist of the same elements, corresponding to a homogeneous Janus structure. Besides, our work suggests that the magnetic ground state of the monolayer is ferromagnetic, with magnetic moments of 1 $\mu_B$ in each $Nb_3$ trimer aligning in-plane (see Figure 1(c) and S5), which agrees well with previous studies[30-32].

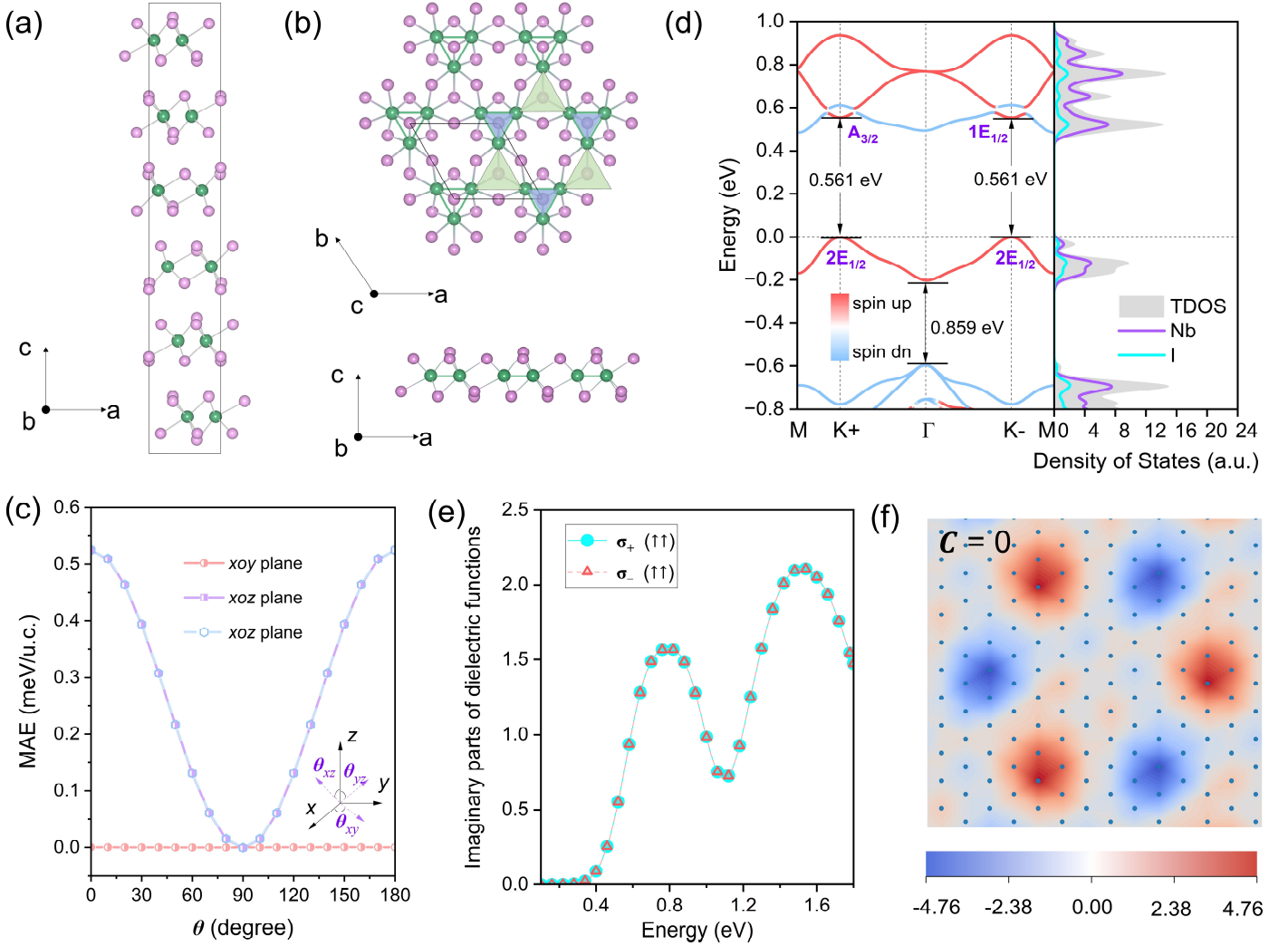

**Figure 1.** A schematic structure for the (a) bulk and (b) monolayer $Nb_3I_8$ from the perspective of side and top views. The green spheres represent Nb atoms and the light purple spheres represent I atoms. (c) The magnetic anisotropy energy (MAE) of monolayer $Nb_3I_8$. The MAE is determined by subtracting the energy of the system in the lowest energy (preferred) magnetic orientation from the energy of the system in other orientations. (d) The band structure considering spin-orbit coupling (SOC) effect and the corresponding element projected density of states (DOS) of $Nb_3I_8$. The irreducible representations (IR) of the valence band maximum and conduction band minimum are also labeled accordingly. (e) The circularly polarized light absorption and (f) corresponding Berry curvatures of the pristine monolayer $Nb_3I_8$. The Chern number is calculated to be 0.

As depicted in Figure 1(d), it is a direct semiconductor with the 0.561 eV bandgap at the K+ and K- high symmetry point. The lowest conduction band (CB) is occupied by spin-up Nb-$d_{xy}/d_{x^2-y^2}$ states, with the irreducible representations (IR) $A_{3/2}$ and $1E_{1/2}$ at K+ and K-, respectively. For the highest valence band (VB) occupied by Nb-$d_{z^2}$, K+ and K- share the same IR of $2E_{1/2}$. According to the great orthogonality theorem, it satisfies the relations: $2E_{1/2} \otimes 1E = A_{3/2}$ and $2E_{1/2} \otimes 2E = 1E_{1/2}$. The initial and final states of the K+ (K-) point can only be excited by left (right) circularly polarized light corresponding to the 1E (2E) symmetry. Such optical selection rules allow K+ and K- to be regarded as valley degrees of freedom. For the pristine Nb$_3$I$_8$ monolayer, the band gaps of two valleys are equivalent, consistent with degenerated absorption excited by left and right circularly polarized light (see Figure 1(e)). The system thus corresponds to a paravalley state. As we expected, the Berry curvature at K+ and K- are equal and opposite in sign, indicating the existence of the valley Hall effect[14].

Normally, the realization of VP is driven by the exchange interaction[21]. Although spin polarization is present here, VP is not yet induced. To understand the underlying mechanisms and develop a general rule for the transition from the paravalley to the VP state, we perform an analytical analysis. The corresponding basis functions for CB and VB are chosen as: $|\psi_c^\tau\rangle = (|d_{x^2-y^2}\rangle - i\tau|d_{xy}\rangle)/\sqrt{2}$ and $|\psi_v^\tau\rangle = |d_{z^2}\rangle$ (More details see Figure S8), where $c$ and $v$ represent CB and VB. $\tau = \pm 1$ is valley index. An effective $k \cdot p$ Hamiltonian is constructed as follows[21]:

$$H(k) = I_2 \otimes H_0(k) + H_{ex}(k) + H_{soc}(k) \tag{1}$$

To reproduce the anisotropic dispersion and more importantly the electron-hole asymmetry, the first term is given by

$$H_0(k) = \begin{bmatrix} \frac{\Delta}{2} + \varepsilon + t'_{11}(q_x^2 + q_y^2) & t_{12}(\tau q_x - iq_y) + t'_{12}(\tau q_x + iq_y)^2 \\ t_{12}(\tau q_x + iq_y) + t'_{12}(\tau q_x - iq_y)^2 & -\frac{\Delta}{2} + \varepsilon + t'_{22}(q_x^2 + q_y^2) \end{bmatrix} \tag{2}$$

where $\Delta$ represents the band gap at the K valleys, $\varepsilon$ denotes a correction energy associated with the Fermi energy, $t$ and $t'$ are the effective nearest-neighbor and second-next-neighbor hopping integral, $q$ is the momentum vector. $I_2$ is the $2 \times 2$ identity matrix.

The second term originates from the intrinsic exchange interaction of Nb-$d$ electrons:

$$H_{ex}(k) = \sigma_z \otimes \begin{bmatrix} -m_c & 0 \\ 0 & -m_v \end{bmatrix} \tag{3}$$

where $\sigma_z$ is the Pauli matrix, $m_c$ ($m_v$) represents the effective exchange splitting in the band edge of CB (VB). The exchange interaction, equivalent to an intrinsic magnetic field, tends to split the spin-majority and spin-minority states.

The most crucial third term, related to spin-orbit coupling (SOC) effect, can be written as

$$H_{soc}(k) = \frac{\tau\lambda}{2} \begin{bmatrix} L_z & L_x - iL_y \\ L_x + iL_y & -L_z \end{bmatrix} \tag{4}$$

Here, $L_x$, $L_y$, $L_z$ are the $2 \times 2$ matrix for $x, y, z$ components of the orbital angular momentum. The perturbation correction from $p$ orbitals of anions and the remote $d_{xz}$ and $d_{yz}$ characters have been ignored. Based on the orbital composition and theoretical derivation[37], the spin splitting induced by the SOC effect is given by $\lambda_c = \alpha cos\theta$ at the top of the conduction band (CB) and $\lambda_v = 0$ at the bottom of the valence band (VB), with $\theta$ representing polar angle in spherical coordinates.

We suppose that the SOC effect is much smaller than the exchange interactions, in consistent to previous[21, 22] and our situation. According to the total Hamiltonian, the band gap near the valleys K± can be easily deduced:

$$\Delta E_\tau = E_{CB}^{\uparrow\uparrow} - E_{VB}^{\uparrow\uparrow} = \Delta + \alpha\tau cos\theta - m_c + m_v \qquad (5)$$

When the magnetic moment points in-plane, i.e., $\theta = 90°$, the two valleys are degenerate, corresponding to the paravalley state of pristine Nb$_3$I$_8$ we discussed before. VP emerges if the magnetic moment has an out-of-plane component $\theta \neq 90°$ and reaches maximum with the magnitude of $2\alpha$ for the case with merely the out-of-plane magnetic moment $\theta = 0°$.

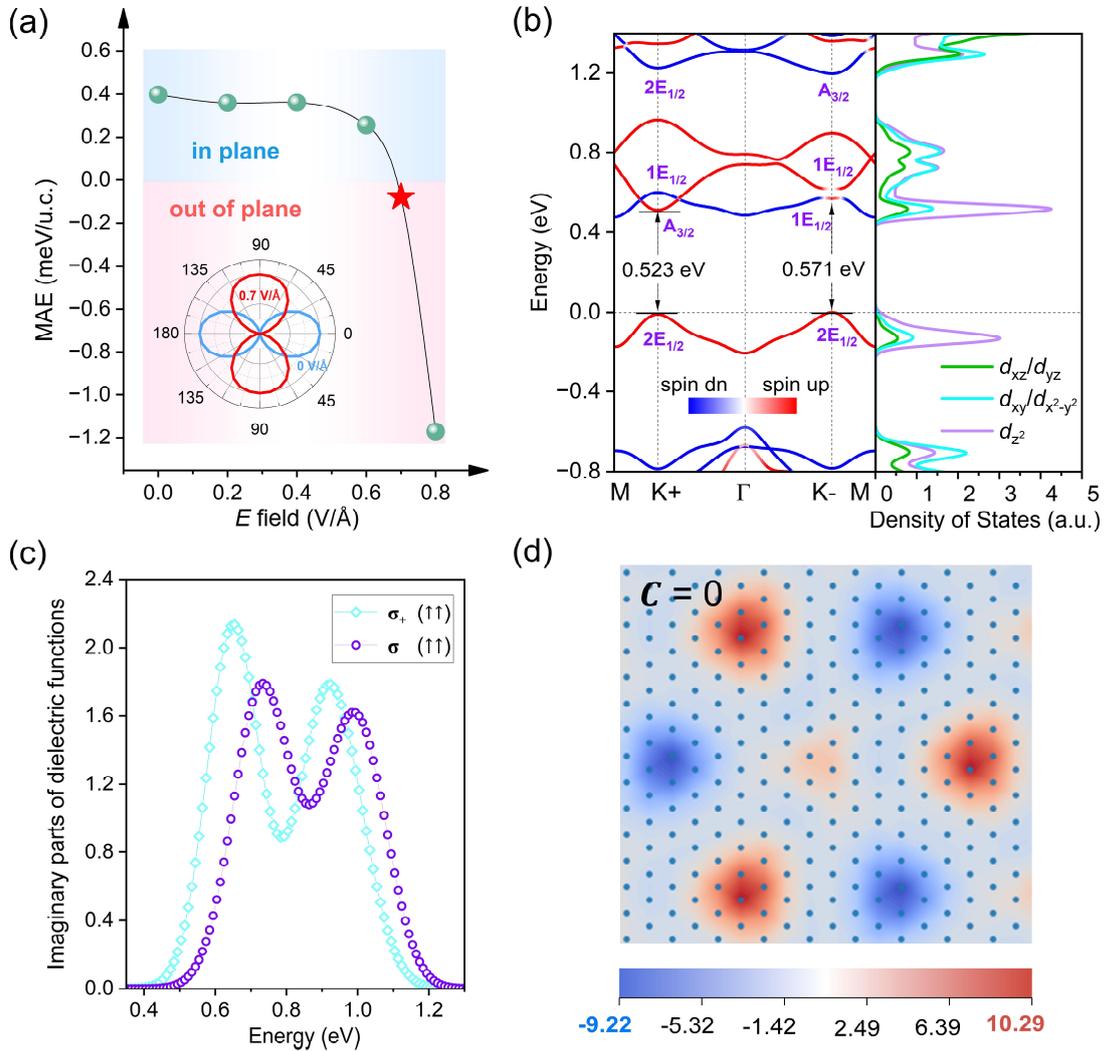

**Figure 2.** (a) The MAE of monolayer Nb$_3$I$_8$ as a function of the applied electric field. The inset shows the variation of normalized MAE in polar coordinates, where the magnetic moment transitions from in-plane (blue) to out-of-plane alignment (red). $\theta$ represents polar angle in spherical coordinates. (b) The band structure and the corresponding projected DOS of Nb element's $d$ orbitals of monolayer Nb$_3$I$_8$ in the VP state. (c) The circularly polarized light absorption and (d) corresponding Berry curvatures of the VP state Nb$_3$I$_8$ monolayer. The Chern number is calculated to be 0.

Following this strategy, VP can be achieved by reorienting the magnetization. Generally, magnetic field is a possible way to do this, yet it's relatively energy-intensive. An electric field, with the advantage of high speed and low energy consumption, is an expected method. The impact of electric field on magnetization is thus explored in Figure 2(a). Its pristine magnetization lies in the plane. If an electric field of ~ 0.7 V/Å is applied, the magnetic moments fully reorient to the $z$-direction. Such impact of electric field on magnetic anisotropy indicates the possibility to induce VP through the magnetoelectric coupling effect. In the case of a fully out-of-plane magnetic moment, as shown in Figure 2(b), the energy band of the K+ valley shifts downward relative to that of the K- valley, creating an energy difference between the two valleys in the conduction band. This energy difference breaks the degeneracy in the absorption of left- and right-circularly polarized light (see Figure 2(c)). Furthermore, the Berry curvature shown in Figure 2(d) differs at the K+ and K- valleys, leading to the existence of the anomalous valley Hall effect. All these evidences demonstrate the emergence of VP state. The VP value reaches up to 48 meV, exceeding the 25 meV threshold for logic operations at room temperature[38]. This is sufficient to withstand thermal noise, indicating the potential for device application. In traditional ferrovalley materials, VP should be controlled by magnetic field, enabling electrically read and magnetically written memory devices[21]. Here, due to the presence of magnetoelectric coupling and AVHE in $Nb_3I_8$, the binary information is stored by the VP that can be controlled by an electric field and read out via the sign of the transverse Hall voltage. All-electric valleytronic devices with energy-efficient advantages are emerging.

In addition to the magnetic properties of $Nb_3I_8$, we observe that its Janus structure inherently exhibits ferroelectricity. The presence of two non-equivalent layers of iodine atoms breaks the spatial inversion symmetry, inducing an out-of-plane electric polarization of 253.8 pC/m, which is consistent with our electrostatic potential calculations shown in Figure S2. More intriguingly, as illustrated in Figures S3 and S6, when the in-plane coordinates are fixed in the breathing-in (where the small triangle of Nb atoms is positioned above the large one in Figure 5(a)) or breathing-out state (the reversed case with the larger triangle situated above), the out-of-plane polarization points upwards or downwards. Due to the coupling between ferroelectricity and structural breathing, the electric field is expected to be an effective way to control the breathing degree of freedom. During the breathing switching process, the ferroelectric polarization reverses from upward to downward with the barrier height of approximately 0.17 eV/atom, comparable to that of $Ti_3X_8$[39-41]. Unlike conventional polarization reversal mechanisms in two-dimensional materials, which typically result from uniform atomic displacements or sliding of atomic layers, this breathing mechanism is characterized by alternating expansions and contractions of the Nb triangular motifs. Given the intrinsic magnetoelectric coupling in the material that governs valley degrees of freedom, we expect that the breathing degree of freedom, coupled with ferroelectricity, will enable control over valley degrees of freedom, introducing a breathing-ferroelectric control mechanism.

To study the impact of such dynamic control on valley properties, we incorporate the breathing degree of freedom ($\kappa$) into the $k \cdot p$ model, where the effect of breathing is similar to the electron exchange interaction described in previous work[22], yielding the following expression derived from equation (5):

$$\Delta E_\tau = E_{CB}^{\uparrow\uparrow} - E_{VB}^{\uparrow\uparrow} = \Delta + \alpha\tau\cos\theta - m_c + m_v + \kappa \quad (6)$$

we find that the band gap of the two valleys can be regulated by the breathing degree of freedom. When the breathing effect decreases the band gap at K+ valley, the one at K− will be reduced as well. Such kind of simultaneous band gap changing provides the possibilities to obtain the desirable critical state of which at one valley the gap is closed while at the other one the gap is still open. Interestingly, for K− valley with $\Delta E_{K-} = 0$, it owns a linear dispersion with a single spin channel.

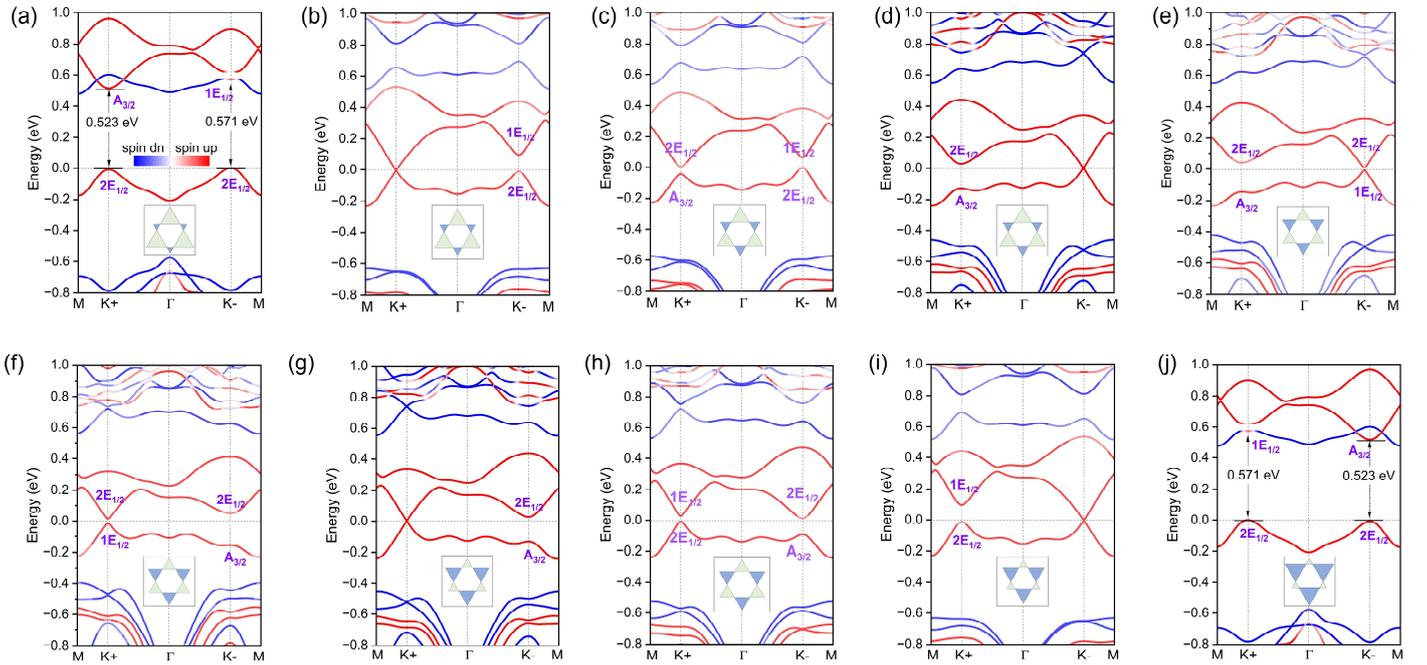

**Figure 3.** The spin-resolved band structure of monolayer Nb$_3$I$_8$ in the process of the breathing ferroelectric switching. (a)-(e) correspond to breathing-in configurations with upward polarization, and (f)-(j) are cases in the breathing-out regime with downward polarization. The irreducible representations are indicated on the corresponding energy bands. The inset illustrates the variation of the Nb framework during the breathing transition, which represents the switching process of ferroelectricity.

First-principles calculations are then carried out to prove the above analysis. As shown in Figure 3(a-h) and Figure S7, when the breathing process starts, the valence band moves upward and the conduction band moves downward. The K+ valley, originally with a smaller band gap, closes first while the bandgap at K- valley still opens, corresponding to the half-valley-metal state (Figure 3(b)). As the breathing progresses, the linear dispersion at K+ reopens, accompanied by a component exchange (Figure 3(c)). In this configuration, the edge states crossing the Fermi level indicate its nontrivial topological characteristics (Figure 4(b)), which can be further demonstrated by a non-zero Chern number $C = 1$. The K+ valley electrons are expected to form protected, quantized conducting channel, demonstrating the emergence of the QAVHE. Following the breathing steps, the K- valley is closed (Figure 3(d)). QAVHE of the new half-valley-metal state transfers from K+ to K- channel. For the case shown in Figure 3(e) with the band gap at K- valley reopened as well, component exchange occurs in both valleys, causing the conductive edge state crossing the Fermi level twice (Figure 4(c)). The Chern numbers at valley K+ and K- are respectively $C_{K+}$ = 1 and $C_{K-}$ = -1. The total Chern number becomes zero, suggesting recovery to the topological trivial VP state. In contrast to the initial state seen in Figure 3(a), VP is reversed here. This scenario demonstrates the possibility of structural breathing to flip VP through electric-field control. When the breathing process continues until it vanishes, the material reaches a non-polarized state (i.e. phase I to phase II in Figure 5(a)). Such a metallic paraelectric state in absence of valley character (Figure S10) is beyond the scope of our discussions. The electric field further reactivates structure into the breathing-out regime (Figure 3(f)-(j)), leading to an out-of-plane ferroelectric state with polarization pointing downward. The transition process from phase II to phase I' exhibits similar characteristics to the breathing-in states with upward polarization (Figure 3(a)-(e)), with the only difference being the exchange of electronic states between the K+ and K- valleys. It is important to note that the symmetry of the K+ and K- valleys also exchanges in accompanying with band dispersion between breathing-in and -out states. The VP in Figure 3(a) and 3(j) is not reversed, as confirmed by their equivalent circularly polarized light absorption (Figure S9).

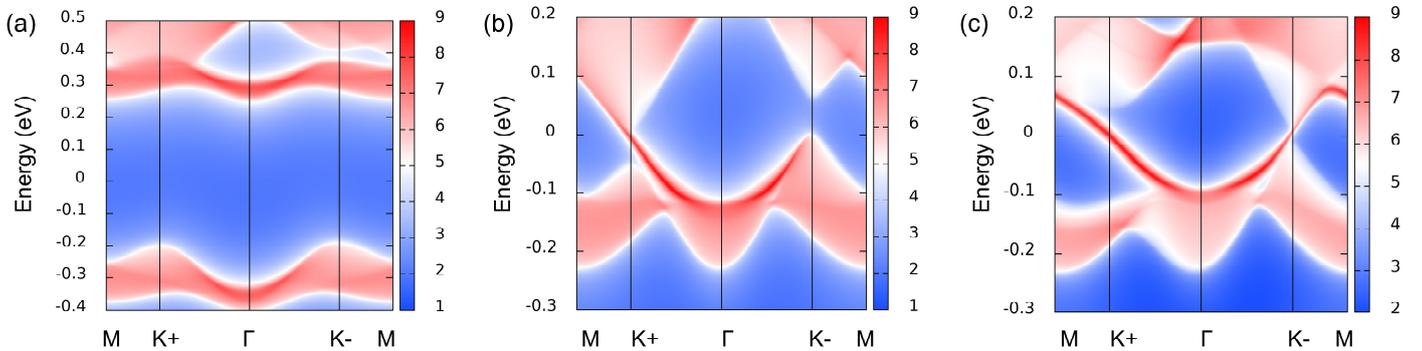

**Figure 4.** During the breathing process, edge states with topological properties emerge at the boundaries of the structure. (a-c) correspond to the electronic states shown in Figure 3(a), (c), (e). The red line represents the conducting edge state, while the light red area indicates the bulk bands.

According to above discussions, we deduce a generalized framework for the impact of breathing on valley degree of freedom in breathing kagome lattices as shown in Figure 5. VP occurs due to the magnetoelectric coupling when an electric field is applied to the pristine paravalley state. Based on the lock-in between structural breathing and ferroelectricity, "breathing-in" and "breathing-out" switching can be further manipulated through electric means. During this lattice dynamical process, VP reversal and the transition from the topologically trivial VP semiconductor to nontrivial half-valley metal states are expected. Within the valley-related states evolution, the transport properties undergo transformations among the valley Hall effect, anomalous valley Hall effect, and quantum anomalous valley Hall effect. Notably, under an in-plane electric field, electrons acquire anomalous velocities transverse to the applied field, leading to electron accumulation with opposite valley indices at opposite edges of the sample. When a vertical electric field induces VP, it creates a valley population imbalance at the edges and generates an additional charge Hall current. Remarkably, upon transitioning to the nontrivial half-valley metal state, topologically protected quantum conduction channels emerge along the sample edges, characterized by fully valley-polarized edge states. The breathing kagome lattices thus serve as a novel platform for realizing all-electric valleytronic devices.

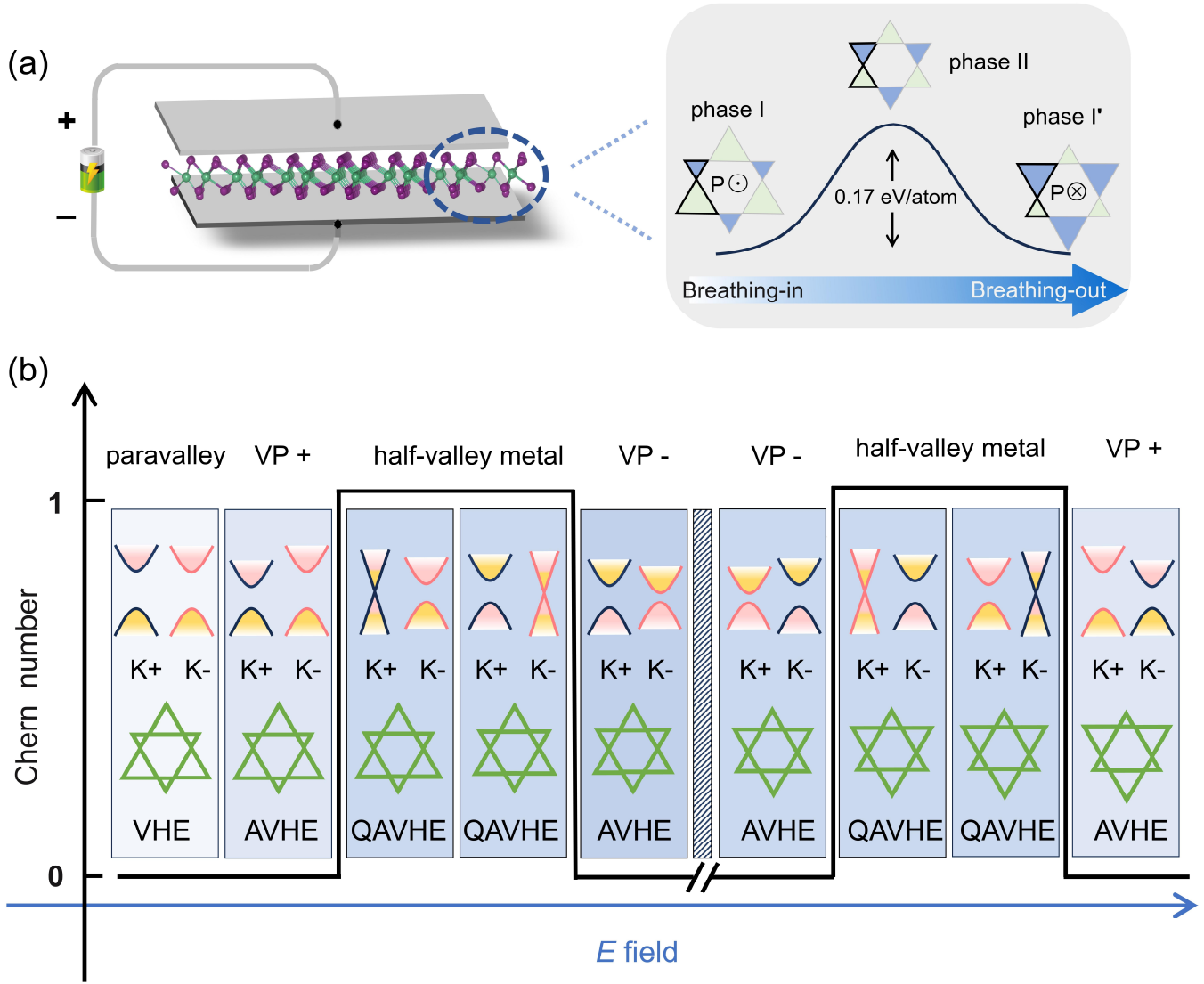

**Figure 5.** (a) Schematic illustration of the electric-field-controlled breathing kagome lattice and the energy barrier associated with its structural breathing transition. (b) Schematic representation of the transition in the electronic band structure during the breathing process, corresponding to distinct valley states. The yellow and pink areas represent the valence and conduction band components, respectively, while the black and red lines denote the K+ and K- valleys. The green framework illustrates the structural changes in the Nb lattice. The paraelectric metallic state, which is beyond the scope of our discussion, is represented by the shaded region.

## CONCLUSION

In conclusion, our study proposes a magnetoelectric coupling mechanism to induce VP in breathing kagome lattices and utilizes the breathing degree of freedom under electric field modulation to achieve VP reversal and multiple valley states, particularly the topological ones with quantum anomalous valley Hall effect. Materials with kagome lattice have recently been regarded as a promising star in the field of condensed matter physics due to their distinctive properties, such as superconductivity and spin/charge density waves. We have further identified prospective research interests of their breathing subfamily for valleytronics. Our study indicates the existence of abundant coupling effects in breathing kagome systems, demonstrating their significance for studies of the fundamental physics in ferroelectronics, spintronics, valleytronics, topology, optics, and their intersecting areas. Given that monolayers with breathing kagome structure have been experimentally synthesized[36] and the interplay between breathing and polarization in such systems has been reported[42], we strongly advocate experimental efforts on valley-related properties within this intriguing family.

**COMPUTATIONAL METHONDS**

The calculations are performed by using the first-principles methods based on the density functional theory The calculations are performed by using the first-principles methods based on the density functional theory within the projector-augmented-wave (PAW) formalism[43], as implemented in the Vienna ab initio simulation package (VASP)[44, 45]. The exchange-correlation functional is treated by the generalized gradient approximation (GGA) approximation within the Perdew-Burke-Ernzerhof (PBE) formalism[46]. The kinetic energy cutoff is set to 500 eV for the plane wave expansion and 5 × 5 × 1 Γ-centred grids are adopted for the first Brillouin zone integral, as others did. The convergence criterion for the electronic energy is $10^{-5}$ eV and the structures are relaxed until the Hellmann–Feynman forces on each atom are less than 1 meV Å. The vacuum space of 15 Å is introduced to avoid interactions between periodically repeated layers. To describe the strong correlation effects of Nb-4$d$ electrons, the GGA+$U$ method is used to describe the on-site Coulomb repulsion between Nb-$d$ electrons[47] and effective $U$ value is chosen to 2 eV according to other works[33, 48]. Berry curvature calculation via available VASPBERRY package[49]. The optical properties are calculated by our own code OPTICPACK[50-52], which has been confirmed to be reliable in optical properties of ferromagnetic materials[21, 22]. The Chern number calculated by the following expression $C = \frac{1}{2\pi}\sum_n \int d^2k \Omega_n$ and edge states are analyzed using the WannierTools software package[53].

**NOTES**

The authors declare no competing financial interest.

**ACKNOWLEDGMENTS**

This work was supported by the National Key Research and Development Program of China (Grants No. 2022YFA1402902 and No. 2021YFA1200700), the National Natural Science Foundation of China (Grants No. 12134003 and No. 12304218), National funded postdoctoral researcher program of China (Grant No. GZC20230809), Shanghai Science and Technology Innovation Action Plan (Grant No. 21JC1402000), Shanghai Pujiang Program (Grant No. 23PJ1402200), and East China Normal University Multifunctional Platform for Innovation.